\input qmass.sty

\rightline{FTUAM-00-07, 2000}
\rightline{hep-ph/xxxxxxxx}
\smallskip
\rightline{\datestamp}
\medskip
\centerline{\bf Improved Determination of the $b$ Quark Mass from Spectroscopy}
\bigskip
\centerline{F. J. Yndur\'ain}
\centerline{Departamento de F\'{\i}sica Te\'orica, C XI}
\centerline{Universidad Aut\'onoma de Madrid, Canto Blanco, E-28049, Madrid}
\centerline{E-mail: fjy@delta.ft.uam.es}
\vskip0.75cm
\setbox0=\vbox{\hsize=14.4cm
\noindent{\petit{\sl Abstract.} Using recently evaluated contributions 
(including a novel one calculated here),
 we present  updated values for the pole mass and $\overline{MS}$ 
mass of the $b$ quark: $m_b=5022\pm58$ MeV, for the pole mass, 
and $\bar{m}_b(\bar{m}_b)=4286\pm36$ MeV for the $\overline{MS}$ one.
These values are accurate including,  
 respectively,   
 $O(\alpha_s^5\log\alpha_s)$  
 and $O(\alpha_s^3)$ corrections 
and, in both cases, leading orders in the ratio $m_c^2/m_b^2$.}}
\centerline{\box0}
\vskip1cm

One of the sources of information for the quark masses is quarkonium spectroscopy. 
By evaluating the $\bar{b}b$ potential including
 relativistic and radiative corrections, 
as well as leading nonperturbative effects,\ref{1,2,3}  
and using this in a perturbative expansion, 
it has been possible to find values of the {\sl pole} quark 
masses with increasing  accuracy;\ref{2,3,4} 
in this note we will go up to 
fourth and leading fifth order, in the approximation of 
neglecting ``light" ($u,\,d,\,s$) quark masses and to leading order
 (actually, $O(\alpha_s m_c^2/m_b^2)$) 
in the $c$ quark mass. 
The connection with the \msbar\ 
mass has been known for some time to one and two loops\ref{5}: very 
recently, a three loop evaluation has been completed. 
Coupling this with the pole mass evaluations, we now have 
an order $\alpha_s^3$ result for the \msbar\ mass. 
We review here briefly this.
\bigskip
\noindent 1. {\sl  $m_b\,-\,\bar{m}_b(\bar{m}_b) $ connection}
\medskip

Write, for a heavy quark,  
$$\bar{m}(\bar{m})\equiv m/\{1+\delta_1+\delta_2+\delta_3+\cdots\};\equn{(1a)}$$
$m$ here denotes the {\sl pole} mass, and $\bar{m}$ is  the \msbar\ one. 
One has
$$\delta_1=C_F\dfrac{\alpha_s(\bar{m})}{\pi},\quad
\delta_2=c_2\left(\dfrac{\alpha_s(\bar{m})}{\pi}\right)^2,\quad
\delta_3=c_3\left(\dfrac{\alpha_s(\bar{m})}{\pi}\right)^3.\equn{(1b)}$$
Here $\alpha_s$ is to be calculated to three loops:
$$\alpha_s(\mu)=\dfrac{4\pi}{\beta_0L}\left\{1-\dfrac{\beta_1\log L}{\beta_0^2L}+
\dfrac{\beta_1^2\log^2L-\beta_1^2\log L+\beta_2\beta_0-\beta_1^2}{\beta_0^4L^2}\right\}
$$
with
$$L=\log\dfrac{\mu^2}{\Lambdav^2};\quad
\beta_0=11-\tfrac{2}{3}n_f,\quad
\beta_1=102-\tfrac{38}{3}n_f,\quad
\beta_2=\tfrac{2857}{2}-\tfrac{5033}{18}n_f+\tfrac{325}{54}n_f^2.$$

The coefficient $c_2$ has been evaluated by Gray et al.\ref{5}, and reads
$$c_2=-K+2C_F,\equn{(1c)}$$
$$\eqalign{K=&\,K_0+\sum_{i=1}^{n_f}\Deltav\left(\dfrac{m_i}{m}\right),\quad
K_0=\tfrac{1}{9}\pi^2\log2+\tfrac{7}{18}\pi^2
-\tfrac{1}{6}\zeta(3)+\tfrac{3673}{288}-\left(\tfrac{1}{18}\pi^2+
\tfrac{71}{144}\right)(n_f+1)\cr
\simeq&\,16.11-1.04\,n_f;\quad 
\Deltav(\rho)=\tfrac{4}{3}\left[\tfrac{1}{8}\pi^2\rho-\tfrac{3}{4}\rho^2+\cdots\right].
 \cr}\equn{(1d)}$$
$m_i$ are the (pole) masses of the quarks strictly lighter 
than $m$, and $n_f$ is the number of these. For the $b$ quark case, $n_f=4$ 
and only the $c$ quark mass has to be considered; we will take $m_c=1.8\,\gev$ 
(see Table 1 below)  
for the calculations.

The coefficient $c_3$ was recently calculated by Melnikov and van~Ritbergen,\ref{6} 
where the exact expression may be found. Neglecting now the $m_i$,
$$c_3\simeq                                  
190.389 - 26.6551 n_f + 0.652694 n_f^2. \equn{(1e)}$$

For the $b$, $c$ quarks, with $\alpha_s$ as given below,
$$\eqalign{\delta_1(b)=&0.090,\cr
\delta_2(b)=&0.045,\cr
\delta_3(b)=&0.029;\cr}\quad
\eqalign{\delta_1(c)=&0.137,\cr
\delta_2(c)=&0.108,\cr
\delta_3(c)=&0.125.\cr}\equn{(2)}$$

From these values we conclude that, for the $c$ quark, 
the series has started to diverge at second order, and it certainly 
diverges at order $\alpha_s^3$. For the $b$ quark
 the series is at the edge of convergence 
for the $\alpha_s^3$ contribution.
 
Take now as input parameters
$$\Lambdav(n_f=4,\,\hbox{three loops})=0.283\pm0.035\;\gev\;
\left[\;\alpha_s(M_Z^2)\simeq0.117\pm0.024\;\right]
$$ 
(ref.~7) and for the gluon condensate, very poorly known, the value 
$\langle\alpha_sG^2\rangle=0.06\pm0.02\;\gev^4.$
From the mass of the $\upsilonv$ particle we 
have a very precise determination for the pole mass of the $b$ quark. 
This determination is correct to order $\alpha_s^4$ 
and including leading $O(m_c^2/m_b^2)$ and 
leading nonperturbative corrections as well as the 
$\alpha_s^5$ corrections proportional to  $\log\alpha_s$; 
the details of it will be given below. With 
 the renormalization point 
$\mu=m_bC_F\alpha_s$ we have,
$$\eqalign{m_b=&
5022\pm43\,(\Lambdav)\;\mp5\,(\langle\alpha_sG^2\rangle)^{-31}_{+37} \;
(\hbox{vary}\; \mu^2\;{\rm by}\,25\%)
\;\pm 38\;({\rm other\; th.\;uncert.})\cr
=&5022\pm58\;\mev.}
\equn{(3a)}$$
Here we append $(\lambdav)$ to the error 
induced by that of $\lambdav$, and likewise 
$(\langle\alpha_sG^2\rangle)$ tags the error due to that of the condensate. 
The error labeled (other th. uncert.) includes also the error evaluated in ref.~8; 
the rest is as in ref.~3.

Using the three loop relation (1) of the pole mass to the \msbar\ mass 
we then find
$$\bar{m}_b(\bar{m}_b)=4284\pm7\;(\lambdav)\mp5\;(\langle\alpha_s G^2\rangle)
\pm35\;(\hbox{other th. uncert.})=4284\pm36\;\mev.\equn{(3b)}$$
The slight dependence 
of $\bar{m}$ on $\lambdav$ when evaluated in this way was already noted in 
ref.~2.

There is another way of obtaining $\bar{m}$, which is to express directly 
the mass of the $\upsilonv$ in terms of it, using
 \equn{(1)} and the order $\alpha_s^3$ formula for the $\upsilonv$ 
mass in terms of the pole mass (see e.g. ref.~2). One finds, for $n_f=4$, 
and neglecting $m_c^2/m_b^2$,
$$M(\upsilonv)=2\bar{m}(\bar{m})
\left\{1+C_F\dfrac{\alpha_s(\bar{m})}{\pi}+7.559\left(\dfrac{\alpha_s(\bar{m})}{\pi}\right)^2
+\left[66.769+18.277\left(\log C_F+
\log\alpha_s(\bar{m})\right)\right]
\left(\dfrac{\alpha_s(\bar{m})}{\pi}\right)^3\right\}.
\equn{(4a)}$$
(One could add the leading nonperturbative contributions to (4a) 
\`a la Leutwyler--Voloshin in the standard way; see e.g. refs.~2,~3,~9).
This method has been at times advertised as improving the 
convergence, allegedly because the \msbar\ mass does not 
suffer from nearby renormalon singularities. 
But a close look to (4a) does not seem to bear this out. 
To an acceptable $O(\alpha_s^4)$ error we can replace 
$\log(\alpha_s(\bar{m}))$ by $\log(\alpha_s(M(\upsilonv/2))$ above.  
With $\lambdav$ as before  
(4a) then becomes
$$M(\upsilonv)=2\bar{m}_b(\bar{m}_b)
\left\{1+C_F\dfrac{\alpha_s(\bar{m})}{\pi}+7.559\left(\dfrac{\alpha_s(\bar{m})}{\pi}\right)^2
+43.502\left(\dfrac{\alpha_s}{\pi}\right)^3\right\}.
\equn{(4b)}$$
This does not look particularly convergent, and is certainly not 
an  
improvement over the expression using the pole mass, where one has 
for the choice\ref{3} $\mu=C_Fm_b\alpha_s$, and 
still neglecting the masses of quarks lighter 
than the $b$, 
$$M(\upsilonv)=2m_b\left\{1-2.193\left(\dfrac{\alpha_s(\mu)}{\pi}\right)^2-
24.725\left(\dfrac{\alpha_s(\mu)}{\pi}\right)^3-
458.28\left(\dfrac{\alpha_s(\mu)}{\pi}\right)^4+
897.93\,[\log\alpha_s] \left(\dfrac{\alpha_s}{\pi}\right)^5\right\}.\equn{(5)}$$
To order three, (5) is actually better\fnote{The convergence of \equn{(5)} 
is still improved if one solves exactly the purely coulombic part of the 
static potential, as was done in refs.~2,~3, where we send for details. 
For example, the $O(\alpha_s^4)$ term becomes
$-232.12(\alpha_s/\pi)^4$.  
This is the method we used to get the values of $m_b$  here.} 
than (4b). What is more, logarithmic 
terms appear in (4) at order $\alpha_s^3$, 
while for the pole mass expression they first show up at $\alpha_s^5$.    
Finally, the direct formula for $M(\upsilonv)$ in terms of the \msbar\ 
mass presents the extra difficulty that the {\sl nonperturbative} 
contribution becomes  larger than than what one has
 for the expression in terms of the pole mass 
($\sim 80$ against $\sim9$ \mev), 
because of the definition of the renormalization point. 
With the purely perturbative expression (4) plus leading 
nonperturbative (gluon condensate) correction one finds the 
value $\bar{m}_b(\bar{m}_b)=4167\mev$, 
rather low. 
\bigskip
\noindent 2. {\sl Improved determination of $m_b$}
\medskip
\equn{(5)} was deduced neglecting the masses of all quarks lighter than the $b$. 
The influence of the nonzero mass of the $c$ quark, the only worth considering, 
will be evaluated now. 
To leading order it only contributes to the $\bar{b}b$ potential 
through a $c$-quark loop in the gluon exchange 
diagram (diagram $f_2$ in ref.~2). The momentum space 
potential generated by a nonzero mass quark through this 
mechanism is then, in the nonrelativistic limit,
$$\widetilde{V}_{c\,{\rm mass}}=-\dfrac{8C_FT_F\alpha_s^2}{{\bf k}^2}
\int_0^1\dd x\,x(1-x)\log\dfrac{m_c^2+x(1-x){\bf k}^2}{\mu^2}.\equn{(6)} $$
We expand in powers of $m_c^2/{\bf k}^2$. The zeroth term 
is already included in (5). The first order correction is
$$\delta_{c\,{\rm mass}}\widetilde{V}=-\dfrac{8C_FT_F\alpha^2_sm_c^2}{{\bf k}^4}. \equn{(7)} $$
In x-space,
$$\delta_{c\,{\rm mass}} V=\dfrac{C_FT_F\alpha_s^2m_c^2}{\pi}\,r.\equn{(8)} $$
This induces the shift in the mass of the $\upsilonv$ of
$$\delta_{c\,{\rm mass}} M(\upsilonv)=\dfrac{3T_F\alpha_s}{\pi}\,
\dfrac{m_c^2}{m_b^2}\,m_b,$$
so \equn{(5)} is modified to
$$\eqalign{M(\upsilonv)=&2m_b\Bigg\{1-2.193\left(\dfrac{\alpha_s(\mu)}{\pi}\right)^2-
24.725\left(\dfrac{\alpha_s(\mu)}{\pi}\right)^3-
458.28\left(\dfrac{\alpha_s(\mu)}{\pi}\right)^4\cr
+&
897.93\,[\log\alpha_s] \left(\dfrac{\alpha_s}{\pi}\right)^5+
\dfrac{3T_F\alpha_s}{2\pi}\,\dfrac{m_c^2}{m_b^2}\Bigg\}.\cr}\equn{(8)}$$
This produces the value quoted in (3a). Note that the (new) correction 
of order $m_c^2/m_b^2$ is responsible for a shift in $m_b$  of
$$\delta_{c\,{\rm mass}}m_b=-35\; \mev,$$ 
substantially larger than the 
$\alpha_s^5\log\alpha_s$ correction 
evaluated by Brambilla et al.\ref{4} which, for the renormalization point 
$\mu=m_bC_F\alpha_s$, gives
$$\delta_{[\alpha_s^5\log\alpha_s]}m_b=
\tfrac{1}{2}m_b[C_F+\tfrac{3}{2}C_A]C^4_F\alpha_s^5(\log\alpha_s)/\pi\simeq-8\mev.$$

We collect in the table the determinations of the $b$ quark mass 
based on spectroscopy, to increasing accuracy. 
The {\sl stability} of the numerical values of the pole mass is 
remarkable: 
the pole masses  all lie within each other error bars.
 The \msbar\ ones show more spread.
\bigskip
\setbox0=\vbox{
\setbox1=\vbox{\offinterlineskip\hrule
\halign{
&\vrule#&\strut\hfil#\hfil&\vrule#&\strut\quad#\quad&\vrule#&\strut\quad#\quad&\vrule#&\strut\quad#\quad&\vrule#&\strut\quad#\cr
 height2mm&\omit&&\omit&&\omit&&\omit&&\omit&\cr 
&\kern0.2em Reference\kern0.2em&&$m_b({\rm pole})$&&$\bar{m}_b(\bar{m}_b^2)$&&$m_c({\rm pole})$&& $\bar{m}_c(\bar{m}_c^2)$& \cr
 height1mm&\omit&&\omit&&\omit&&\omit&&\omit &\cr
\noalign{\hrule} 
height1mm&\omit&&\omit&&\omit&&\omit&&\omit&\cr
&TY&& $4971\pm72$&&$4401^{+21}_{-35}$\vphantom{$4^{4^4}_{4_4}$}&&
$1585\pm 20\,(^*)$&&$1321\pm 30\,(^*)$\phantom{\big{]}}&\cr
&PY&& $5065\pm60$&&$4455^{+45}_{-29}$\vphantom{$4^{4^4}_{4_4}$}&&
$1866^{+215}_{-133}$&&$1542^{+163}_{-104}$&\cr
&Here&& $5022\pm58$&&$4286\pm36$&&
$-$&&$-$\vphantom{\big\}}&\cr
 height1mm&\omit&&\omit&&\omit&&\omit&&\omit&\cr
\noalign{\hrule}}
\vskip.05cm}
\centerline{\box1}
{\petit
\centerline{{\bf Table 1.} $b$ and $c$ quark masses.\quad $(^*)$
 Systematic errors not included.}}
\vskip-0.2cm
\centerrule{0.3cm}
\smallskip
\setbox2=\vbox{\hsize=0.95\hsize 
\petit{
\noindent
TY: Titard and Yndur\'ain\ref{2}. $O(\alpha_s^3)$ plus $O(\alpha_s^3)v$, $O(v^2)$ 
for $m$;
$O(\alpha_s^2)$ for $\bar{m}$. 
 Rescaled for 
$\lambdav(n_f=4)=283\,\mev$.\hb
PY: Pineda and Yndur\'ain\ref{3}. Full $O(\alpha_s^4)$ for $m$; 
$O(\alpha_s^2)$ for $\bar{m}$. Rescaled for 
$\lambdav(n_f=4)=283\,\mev$.\hb
Here: This calculation. $O(\alpha_s^4)$, $O(\alpha_s m_c^2/m_b^2)$ 
and $O(\alpha_s^5\log\alpha_s)$ 
 for $m$; $O(\alpha_s^3)$ and  $O(\alpha_s^2 m_c^2/m_b^2)$  for  $\bar{m}$. 
Values not given for the $c$ 
quark, as the higher order terms are as large as the leading ones.}}
\centerline{\box2}}
\centerline{\box0}
\centerrule{0.3cm}
\smallskip
We finally remark that the values of $m_b$ quoted e.g. in 
the Table~1 were {\sl not} obtained solving \equn{(8)}, but 
solving exactly the coulombic part of the interaction,
 and perturbing the result (see refs.~2,~3 for details). 
We also note that, in the determinations of $m_b$, the new pieces,
 $O(\alpha_s m_c^2/m_b^2)$ 
and $O(\alpha_s^5\log\alpha_s)$, have been evaluated to first order; 
in particular, we have 
included the corresponding shifts in the {\sl central} values, not in the errors. 
If we included these, the errors would decrease by some 7\%.
 
\bigskip
\noindent{\sl Acknowledgements}. I am grateful to Dr. T.~van~Ritbergen 
for encouragement and 
discussions. Thanks are due to CICYT, Spain, for financial support.
\bigskip
\noindent{\sl References}
\smallskip
\item{1.}{W. Fischler, {\it Nucl. Phys.} B {\bf 129}, 157 (1977); 
A.~Billoire, {\it Phys. Lett.} B {\bf 92}, 343 (1980); 
S. N. Gupta and S. Radford, {\it Phys. Rev.} D {\bf 24},  2309 (1981) and (E) 
D {\bf 25},  3430 (1982); S.~N.~Gupta, S.~F.~Radford 
and W.~W.~Repko, ibid D {\bf 26},  3305 (1982); 
M. Peter, {\it Phys. Rev. Lett.} {\bf 78}, 602 (1997); 
Y.~Schr\"oder, {\it Phys. Lett.} {\bf B447}, 321 (1999).}
\item{2.}{S. Titard and F. J. Yndur\'ain, {\it Phys. Rev.} D {\bf 49}, 6007 (1994) 
and D {\bf 51}, 6348 (1995).}
\item{3.}{A. Pineda and F. J. Yndur\'ain, {\it Phys. Rev.} D {\bf 58}, 3003 (1998), 
and hep-ph/9812371, 1998, in press in Phys. Rev.}
\item{4.}{N.~Brambilla et al., {\it Phys. Lett.} B {\bf 470}, 215 (1999).}
\item{5.}{R. Coquereaux, {\it Phys. Rev.} D {\bf 23}, 1365 (1981); R. Tarrach, 
{\it Nucl. Phys.} B {\bf 183}, 384 (1981); N.~Gray et al., {\it Z. Phys.} C {\bf 48}, 673 (1990).}
\item{6.}{K.~Melnikov and T.~van~Ritbergen, hep-ph/9912391.}
\item{7.}{See J.~Santiago and F.~J.~Yndur\'ain, 
{\it Nucl. Phys.} B {\bf 563}, 45 (1999), and work quoted there.}
\item{8.}{W. Lucha and F. F. Sch\"oberl, UWThPh-1999-77 (hep-ph/0001191).}
\item{9. }{M. B. Voloshin, {\it Nucl. Phys.} B {\bf 154},  365 (1979) and 
{\it Sov. J. Nucl. Phys.} {\bf 36}, 143  (1982); 
H.~Leutwyler, {\it Phys. Lett.} B {\bf 98},  447 (1981).}

\bye